\documentclass[twocolumn,twoside,slac_two]{revtex4}
\usepackage{graphicx}
\usepackage{fancyhdr}
\def\amg{{$\alpha_{\mu \gamma}$~}}

\usepackage{aas_macros}

\begin{document}

\title{Study of microwave$/$gamma-ray properties for $Fermi$-LAT bright AGNs}

%

\author{D. Gasparrini, E. Cavazzuti, P. Giommi, C. Pittori,  S. Colafrancesco on behalf  of  Fermi LAT  collaboration}
\author{}
\affiliation{Agenzia Spaziale Italiana (ASI) Science Data Center, I-00044 Frascati (Roma), Italy}

\begin{abstract}
Blazars are a small fraction of all extragalactic sources but, unlike other objects, they are strong emitters across the entire electromagnetic spectrum. Recent data in the microwave region of the electromagnetic spectrum have become available to allow for systematic studies of blazars over large cosmological volumes. This frequency band is indeed particularly suited for the selection of blazars since at these frequencies the contamination from radio extended components with steep spectra is no longer present and the emission from the accretion process is negligible. During the first 3 months of scientific operations Fermi-LAT detected 106 bright, high-galactic latitude ($| b |> 10$ deg) AGNs with high significance. In this study we investigate the possible relations between the microwave and the gamma-ray emissions for $Fermi$-LAT detected AGNs belonging to WMAP 5th year bright source catalog.
\end{abstract}

\maketitle

\thispagestyle{fancy}


\section{Introduction}
Blazars are rare extragalactic objects as they are a subset of radio loud QSOs, which in turn are only $\approx $10\% of radio-quiet QSOs and Seyfert galaxies that are found in large numbers at optical and at X-ray frequencies. Despite that, the strong emission at all wavelengths that characterizes blazars, makes them the dominant type of extragalactic sources in those energy windows where the accretion onto a supermassive black hole, or other thermal mechanisms, do not produce significant radiation. For instance, in the microwave band, \cite{giommiWMAP3} showed that blazars are the largest population of extragalactic objects (see also \citet{toffolatti98}). The same is true in the $\gamma$-ray band  \citep{AbdoAGNpaper} and at TeV energies where BL Lac objects are the most frequent type of sources found in the high Galactic latitude sky.
 
 Blazars have been known and studied in different energy windows for over 40 years, however, many questions still remain open about their physics and demographics,  e.g., how are jets made and how are they accelerated? How are the relativistic particles accelerated and why is the maximum acceleration so much higher in BL Lacs? What are the mechanisms producing  blazar variability and what is the blazar duty-cycle? 
 
 We investigate these problematics using a multi-wavelength approach through WMAP ($\mu$-wave) and Fermi ($\gamma$-ray) observations.
 
Primordial photons are redshifted to $\mu$-wave frequencies due to the Universe expansion and we see these photons as cosmic background in $\mu$-wave band. 
Tiny inhomogeneities in the early universe left their imprint on the CMB in the form of small anisotropies in its temperature. These anisotropies contain information about basic cosmological parameters (e.g. total energy density and curvature of the universe). The ability of the present and upcoming CMB experiments to determine the cosmological parameters requires a careful cleaning of the CMB maps from the galactic and extra-galactic foregrounds . This cleaning produces very useful bright source catalogs useful to our scopes.

The microwave frequency range is, moreover, likely the best region of the electromagnetic spectrum
to pursue statistical studies of blazars since it is least affected by the superposition
of spectral components of different nature like e.g. steep radio emission from the extended
part of the jet, non-nuclear optical emission from the host galaxy or optical/UV and soft
X-ray emission from an accretion disk.

We have already studied a $\mu$-wave selected sample and its X-ray properties \citep{giommiSwift} and verified that the $\mu$-wave flux is tightly correlated with the X-ray one. Given the currently accepted emission mechanisms, such as Synchrotron Self Compton, as responsible for the broad-band blazar emission, we would like to investigate a possible relation among $\mu$-wave and $\gamma$-ray fluxes.  

 Our strategy consists in building a complete sample of blazars selected in the $\mu$-wave band, starting from the WMAP bright source catalogues periodically released by the WMAP Team. Than we study the properties of this $\mu$-wave selected sample of jet-dominated AGN looking for synchrotron peak distribution of the sources, $\gamma$-ray spectral index estimates, possible relations between $\mu$-wave fluxes and X-ray, $\gamma$-ray duty cycle, \amg. As in \citet{pittori07}, we take into account
the constraints from the
observed extragalactic
gamma-ray background on
the maximum duty cycle allowed
for the selected sample of WMAP Blazars.

\section{Instruments}

Wilkinson Microwave Anisotropic Prope (WMAP) \citep{bennett03} satellite provides an all-sky survey of the millimeter-wave sky and its point source catalogs are valuable for the study of flat-spectrum radio sources. In this work we use the five-year catalogue of 390 foreground sources (WMAP5) \citep{wright09}. We define the complete sample as including all sources in the WMAP5 catalogue with flux at 41 GHz larger than 0.9 Jy.
 We define a flux limited sample using the source fluxes as observed in the WMAP 41 GHz channel to have a compromise between sensitivity, completeness and the need to use a high frequency band, to avoid the steep radio components. Highest frequency channels in WMAP survey (61, 94 GHz) have less objects and are less deep in flux with respect to lower frequencies (23, 33, 41 GHz). 

$Fermi$ Gamma-ray Space Telescope is an international mission suited for the study of $\gamma$-ray emission in the Universe. The LAT, the primary instrument, is an imaging, wide field-of-view telescope, covering the energy range from below 20 MeV to more than 300 GeV with a sensitivity that exceeds EGRET by a factor of 30 or more.
$Fermi$ satellite commonly observes the sky in scanning mode: on a given orbit, the LAT will sweep the sky 35 degree away from the orbital plane, covering 75\% of the sky.  At the end of the orbit, $Fermi$ will rock to 35 degree on the other side of the orbital. Therefore the entire sky is covered every three hours providing an uniform sky coverage within few orbit \citep{atwood09}. Thanks to the improved instrument performance (point spread function, effective area, large field of view, broad energy range) blazar observations with Fermi LAT are advancing our understanding of blazars and AGNs.

 \section{WMAP 5th year and $Fermi$ samples}

As in \citet{giommiWMAP3}, we have carried out an extensive search to identify the counterparts of all the microwave foreground sources listed in the WMAP 5-year catalogue using literature and archival data. Our results allow us to define a flux limited sample of 254 high Galactic latitude microwave sources ($f_{41GHz} \ge 0.9$ Jy, $|b_{\rm II}| > 15^\circ$) which is virtually complete. From the analysis of classification in this complete sample we note that there is a dominance of blazars (~83\%).
Most of these are Low Synchrotron Peaked blazars (LSP, mainly FSRQ, ~80\% of the total). This is what we expect from a sample selected in microwave band, in fact High Synchrotron Peaked Blazars (HSP) are fainter than LSP at millimetric wavelengths.
 
\begin{figure}
\center{
\includegraphics[height=7cm,angle=-90]{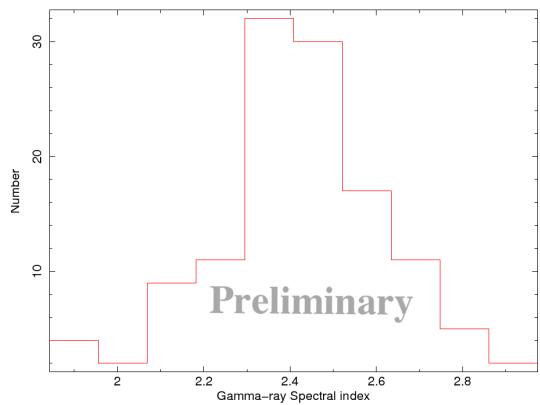}
\caption{Photon index distribution for the WMAP-$Fermi$-LAT sources}}
\label{spindex}
\end{figure}

 We perform a spatial cross-correlation between the WMAP5 sample and the $Fermi$ Bright Source List \cite{AbdoLATpaper}, including also the low-significance (more than 5 sigma) sources.
In this way we obtain 69 objects with known microwave emission that are detected in the early phase of Fermi data acquisition. As expected 95\% of this sample are blazars, all objects well known from literature. 
 
We compare the distributions of $\gamma$-ray properties such as spectral index and flux: the spectral index is centered around 2.4 (fig. \ref{spindex}) and for most of the objects it is above 2, according to the FSRQ spectral index distribution as in \citet{AbdoAGNpaper} and the dominance of FSRQs in WMAP5 sample.

As in \citet{pittori07}, we define a $\mu$-wave to $\gamma$-ray slope as
\begin{equation}
 \displaystyle {\alpha_{\mu \gamma} = -{Log(f_{94GHz}/f_{100MeV})\over{Log(\nu_{94GHz}/\nu_{100MeV})}}}~,
 \label{eq.alphamugamma}
\end{equation}
and a limiting value: ${\alpha_{\mu \gamma}}_{100\% CGB} = 0.994$ which is the
value of an hypotetical source that would produce 100\% of the CGB if
representative of the class.\\
Any source with $\alpha_{\mu \gamma} < $ 0.994 should have a duty cycle lower
than 100\% in order not to overproduce the extragalactic diffuse $\gamma$-ray
background.\\
We estimate the blazar duty cycle by defining
\begin{equation}
 \displaystyle {Duty ~ Cycle =100 \times 10^{-11.41 ~(0.994-\alpha_{\mu \gamma})}}~,
 \label{eq.dc}
\end{equation}
where $Log(\nu_{94GHz}/\nu_{100MeV}) = 11.41$.

Looking at the first 3 months of data, we have found mostly flaring blazars like EGRET and we did not found strong differences in \amg for the two blazar classes, FSRQ and BL Lac.

\begin{figure} 
\center{
\includegraphics[height=7cm]{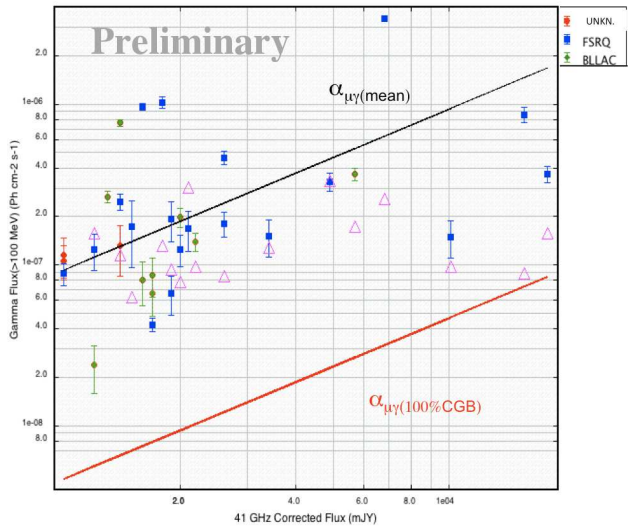}
\caption{$\gamma$-ray Flux ($>$100MeV) vs $\mu$-wave flux density at 41GHz. Blue, green and red points are FSRQ, BLLac and BZU sources, respectively, common to both WMAP5 and $Fermi$ BSL. Magenta triangles are previous EGRET detections, in order to show for variability. The black and red lines show the $\alpha_{\mu \gamma}$ mean and limit respectively.}}
\label{amugamma}
\end{figure}

Moreover, relating the $\gamma$-ray flux (E$>$100  MeV) with the $\mu$-wave flux density at 41GHz, apparently there is not a strong correlation (see fig. \ref{amugamma}). However, in a recent paper \citet{kovalev09}  found a correlation between high frequency radio (15 GHz) and $\gamma$-ray  fluxes. In this Kovalev's work the set of radio observations were quasi simultaneous to $Fermi$ $\gamma$-ray fluxes. We are considering the possibility that the $\gamma$-ray high variability of blazars blurs the correlation with the $\mu$-wave band.

\begin{figure}
\center{
\includegraphics[height=7cm,angle=-90]{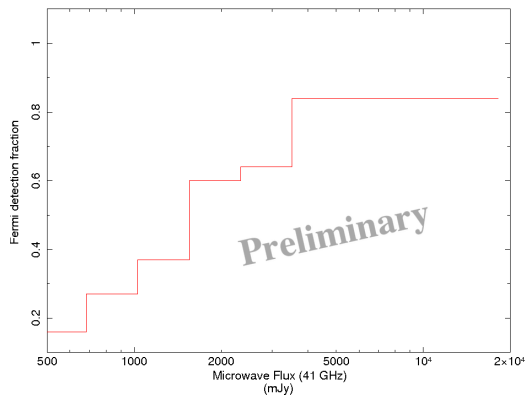}
\caption{Fraction of the $Fermi$-LAT detected sources wrt $\mu$-wave flux.}}
\label{fraction}
\end{figure}

If we bin the microwave flux density, we can see that the fraction of $Fermi$-LAT detected sources increases for brightest $\mu$-wave emitters (see fig. \ref{fraction}). This is another hint of a relation between emission processes in microwave and gamma band.

\section{Conclusions}

In our preliminary analysis we find that
\begin{itemize}

\item Fermi is detecting a significant number (although a small fraction) of powerful microwave emitters. 

\item The $\mu$-wave to X-ray correlation is tight.

\item The $\mu$-wave to $\gamma$-ray correlation is blurred by large $\gamma$-ray variability of blazars, specially with non-simultaneous data.
\end{itemize}
Planck data along with $Fermi$ one, being both operated in survey mode, will give us unprecedented set of real simultaneous data.
 
These two energy bands ($\mu$-waves and $\gamma$-rays) are indeed best suited to study
blazars as their emission at these frequencies is largely dominated by non-thermal
radiation. The combined study of blazars in the $\mu$-wave and in the $\gamma$-ray energy
bands will probably offer a unique opportunity to understand many of the physical details of
this class of black hole-dominated cosmic structures.
\\

\begin{acknowledgments}
The $Fermi$ LAT Collaboration acknowledges support from a number of agencies and institutes for both development and the operation of the LAT as well as scientific data analysis. These include NASA and DOE in the United States, CEA/Irfu and IN2P3/CNRS in France, ASI and INFN in Italy, MEXT, KEK, and JAXA in Japan, and the K.~A.~Wallenberg Foundation, the Swedish Research Council and the National Space Board in Sweden. Additional support from INAF in Italy and CNES in France for science analysis during the operations phase is also gratefully acknowledged.
We acknowledge the use of data and software facilities from the ASI
Science Data Center (ASDC), managed by the Italian Space Agency (ASI).  Part of this work is based on 
archival data and on bibliographic information obtained from the NASA/IPAC Extragalactic Database (NED) and
from the Astrophysics Data System (ADS).
\end{acknowledgments}

\bigskip 
\bibliography{Proceedings_Gasparrini}

\begin{thebibliography}{10}
\expandafter\ifx\csname natexlab\endcsname\relax\def\natexlab#1{#1}\fi
\expandafter\ifx\csname bibnamefont\endcsname\relax
  \def\bibnamefont#1{#1}\fi
\expandafter\ifx\csname bibfnamefont\endcsname\relax
  \def\bibfnamefont#1{#1}\fi
\expandafter\ifx\csname citenamefont\endcsname\relax
  \def\citenamefont#1{#1}\fi
\expandafter\ifx\csname url\endcsname\relax
  \def\url#1{\texttt{#1}}\fi
\expandafter\ifx\csname urlprefix\endcsname\relax\def\urlprefix{URL }\fi
\providecommand{\bibinfo}[2]{#2}
\providecommand{\eprint}[2][]{\url{#2}}

\bibitem[{\citenamefont{{Giommi} et~al.}(2009)\citenamefont{{Giommi},
  {Colafrancesco}, {Padovani}, {Gasparrini}, {Cavazzuti}, and
  {Cutini}}}]{giommiWMAP3}
\bibinfo{author}{\bibfnamefont{P.}~\bibnamefont{{Giommi}}},
  \bibinfo{author}{\bibfnamefont{S.}~\bibnamefont{{Colafrancesco}}},
  \bibinfo{author}{\bibfnamefont{P.}~\bibnamefont{{Padovani}}},
  \bibinfo{author}{\bibfnamefont{D.}~\bibnamefont{{Gasparrini}}},
  \bibinfo{author}{\bibfnamefont{E.}~\bibnamefont{{Cavazzuti}}},
  \bibnamefont{and} \bibinfo{author}{\bibfnamefont{S.}~\bibnamefont{{Cutini}}},
  \bibinfo{journal}{\aap} \textbf{\bibinfo{volume}{508}}, \bibinfo{pages}{107}
  (\bibinfo{year}{2009}).

\bibitem[{\citenamefont{{Toffolatti} et~al.}(1998)}]{toffolatti98}
\bibinfo{author}{\bibfnamefont{L.}~\bibnamefont{{Toffolatti}}}
  \bibnamefont{et~al.}, \bibinfo{journal}{\mnras}
  \textbf{\bibinfo{volume}{297}}, \bibinfo{pages}{117} (\bibinfo{year}{1998}).

\bibitem[{\citenamefont{{Abdo} et~al.}(2009{\natexlab{a}})}]{AbdoAGNpaper}
\bibinfo{author}{\bibfnamefont{A.~A.} \bibnamefont{{Abdo}}}
  \bibnamefont{et~al.}, \bibinfo{journal}{\apj} \textbf{\bibinfo{volume}{700}},
  \bibinfo{pages}{597} (\bibinfo{year}{2009}{\natexlab{a}}).

\bibitem[{\citenamefont{{Giommi} et~al.}(2007)}]{giommiSwift}
\bibinfo{author}{\bibfnamefont{P.}~\bibnamefont{{Giommi}}}
  \bibnamefont{et~al.}, \bibinfo{journal}{\aap} \textbf{\bibinfo{volume}{468}},
  \bibinfo{pages}{571} (\bibinfo{year}{2007}).

\bibitem[{\citenamefont{{Pittori} et~al.}(2007)}]{pittori07}
\bibinfo{author}{\bibfnamefont{C.}~\bibnamefont{{Pittori}}}
  \bibnamefont{et~al.}, \bibinfo{journal}{\apss}
  \textbf{\bibinfo{volume}{309}}, \bibinfo{pages}{89} (\bibinfo{year}{2007}).

\bibitem[{\citenamefont{{Bennett} et~al.}(2003)\citenamefont{{Bennett}, {Hill},
  {Hinshaw} et~al.}}]{bennett03}
\bibinfo{author}{\bibfnamefont{C.~L.} \bibnamefont{{Bennett}}},
  \bibinfo{author}{\bibfnamefont{R.~S.} \bibnamefont{{Hill}}},
  \bibinfo{author}{\bibfnamefont{G.}~\bibnamefont{{Hinshaw}}},
  \bibnamefont{et~al.}, \bibinfo{journal}{\apjs}
  \textbf{\bibinfo{volume}{148}}, \bibinfo{pages}{97} (\bibinfo{year}{2003}).

\bibitem[{\citenamefont{{Wright} et~al.}(2009)\citenamefont{{Wright}, {Chen},
  {Odegard} et~al.}}]{wright09}
\bibinfo{author}{\bibfnamefont{E.~L.} \bibnamefont{{Wright}}},
  \bibinfo{author}{\bibfnamefont{X.}~\bibnamefont{{Chen}}},
  \bibinfo{author}{\bibfnamefont{N.}~\bibnamefont{{Odegard}}},
  \bibnamefont{et~al.}, \bibinfo{journal}{\apjs}
  \textbf{\bibinfo{volume}{180}}, \bibinfo{pages}{283} (\bibinfo{year}{2009}).

\bibitem[{\citenamefont{{Atwood} et~al.}(2009)}]{atwood09}
\bibinfo{author}{\bibfnamefont{W.~B.} \bibnamefont{{Atwood}}}
  \bibnamefont{et~al.}, \bibinfo{journal}{\apj} \textbf{\bibinfo{volume}{697}},
  \bibinfo{pages}{1071} (\bibinfo{year}{2009}).

\bibitem[{\citenamefont{{Abdo} et~al.}(2009{\natexlab{b}})}]{AbdoLATpaper}
\bibinfo{author}{\bibfnamefont{A.~A.} \bibnamefont{{Abdo}}}
  \bibnamefont{et~al.}, \bibinfo{journal}{\apjs}
  \textbf{\bibinfo{volume}{183}}, \bibinfo{pages}{46}
  (\bibinfo{year}{2009}{\natexlab{b}}).

\bibitem[{\citenamefont{{Kovalev} et~al.}(2009)\citenamefont{{Kovalev},
  {Aller}, {Aller} et~al.}}]{kovalev09}
\bibinfo{author}{\bibfnamefont{Y.~Y.} \bibnamefont{{Kovalev}}},
  \bibinfo{author}{\bibfnamefont{H.~D.} \bibnamefont{{Aller}}},
  \bibinfo{author}{\bibfnamefont{M.~F.} \bibnamefont{{Aller}}},
  \bibnamefont{et~al.}, \bibinfo{journal}{\apjl}
  \textbf{\bibinfo{volume}{696}}, \bibinfo{pages}{L17} (\bibinfo{year}{2009}).

\end{thebibliography}






\end{document}